\documentclass{article}
\usepackage{ijcai18}

\usepackage{times}
\usepackage{xcolor}
\usepackage{soul}
\usepackage[utf8]{inputenc}
\usepackage[small]{caption}
\usepackage{graphicx}
\title{``AIded with emotions'' -- a new design approach \\ towards affective computer systems%
\thanks{This work was presented at 1st Workshop on Humanizing AI (HAI) at IJCAI'18 in Stockholm, Sweden.}}

\author{
Barbara Giżycka$^1$, 
Grzegorz J. Nalepa$^1$, 
Paweł Jemioło$^1$, 
\\ 
$^1$ AGH Univeristy of Science and Technology,\\
Al. Mickiewicza 30, 30-059 Krakow, Poland
\\
bgizycka@agh.edu.pl, gjn@agh.edu.pl, 
paweljemiolo@gmail.com \\ 
\vspace{0.5cm}
}
\begin{document}
	
	\maketitle
	
	\begin{abstract}
          As technologies become more and more pervasive, there is a need for %such systems to
          considering the affective dimension of interaction with computer systems to make them more human-like. 
		Current demands for this matter include accurate emotion recognition, reliable emotion modeling, and use of unobtrusive, easily accessible and preferably wearable measurement devices. 
		While AI methods provide many possibilities for better affective information processing, it is not a~common scenario for both emotion recognition and modeling to be integrated in the design phase. 
		To address this concern, we propose a new approach based on affective design patterns in the context of video games, together with summary of experiments conducted to test the preliminary hypotheses. %so far to test our suggestions. 
%		We firmly believe that our approach can be applied to other areas of affective AI applications. 
	\end{abstract}
	
	\section{Introduction}
	
	People nowadays deal with computers and other computer-based devices on everyday basis. 
	It has been several decades since technologies ceased to be used just by a narrow group of users (such as engineers or academia workers). 
	However, casual users have different needs and demands regarding the appliances they interact with. 
	Supporting this process to make it as natural and easy as possible becomes a necessary aspect to consider when developing new systems and applications. 
	
	New branches of studies emerged that inquire exactly the problems and solutions for bringing people's encounters with machines to a friendlier level. 
	Human-Computer-Interaction (HCI), together with Affective Computing (AfC), are one of such interdisciplinary domains. 
	Their origins can be traced back to 1980s and 1990s respectively, and both benefit from findings of computer science, psychology, anthropology and ergonomics, among others. 
	HCI embraces a vast range of interaction aspects, including interface design, controls, and usability engineering. 
	
	Essentially, \textit{usability} is a key concept of HCI. 
	Originally it was equated to the degree of simplicity concerning the learning to use the system, using and maintaining it, and it's infallibility. 
	Today, it has grown to encompass more characteristics, such as fun factor, as well as efficiency and creativity enhancement. 
	Each of these features is influenced by the affective dimension of interaction, meaning that the general emotional state of the user is of crucial importance. 
	As such, this essentially means that human emotions in human-computer interaction have to be taken into consideration~\cite{thompson2015affhci}. 
	
	It was year 1997, when Rosalind Picard from MIT Lab published ``Affective Computing'', a first handbook of the discipline~\cite{picard1997affective}. 
	Sometimes referred to as emotion AI, this approach aspires to collect data of human physiological states, as well as information on behavior metrics, and use this data to develop affect models of the user. 
	Later, these can be used to infer on user's affective responses to what is happening during interaction with the system, and make the software behave just like it ``understood'' human emotions. 
	The motivation for such an attitude is to make computer systems more human-like, by means of augmenting the interaction with them with the affective dimension. 
	What is more, this provides an answer for the need of personalized systems, capable of adapting to the individual preferences and habits of specific users. 
	Selected fields of application of AfC include % range from AI dependent software and
        context-aware recommendation systems, tutoring applications, therapy and telemedicine, or video games and serious games.

	%%% bold statement with no arguments
	%%To fulfill AfC objectives, AI methods seem to be the only satisfying approach.
	In affective computer systems, large amounts of sensor data need to be recorded and processed.
	Therefore, firstly, various AI techniques, mainly machine learning and probabilistic graphical models, provide effective means to analyze such immense information, and derive meaningful interpretations of it~\cite{KollerFriedman:09}. 
%	By combining, for example, artificial neural networks or Markov decision processes with affect detection and interpretation, computer's behavior can appear more complex and intelligent, yet natural. 
	Additionally, certain methods, such as symbolic reasoning, can make the whole system more understandable for the user, thus improving it's transparency and implied controllability.
	
	In our research, we set out from the context of applying AfC solutions in the field of video games.
	We reach for the higher perspective, and aim to show how integration of affect detection and affect modeling can be handled early in the design phase.
	Furthermore, we emphasize the importance of using AI methods for developing necessary data processing and emotion modeling modes.
	
	This paper's original contribution is brought by engaging the affective loop mechanism in a way that is introduced on the design level.
	To ensure dynamic human-computer interaction that takes emotional dimension into the account, we suggest a design proposition based on affective game design patterns. 
	We argue that physiological reaction patterns, corresponding to emotional responses of the person, can be identified and correlated with the affective design patterns~\cite{bgc2018gem}. 
	Notably, we presume that the mode of affectively meaningful human-computer interaction can be conducted with use of unobtrusive, wearable sensory devices~\cite{gjn2017fgcs}. 
	
	All things considered, we propose a new approach to designing affective video games, which enables integration of emotion detection and emotion modeling. 
	We believe that it is important and novel that our design concept allows for these aspects to meet already in the design phase. 
	At the same time, we enunciate the benefits of using AI methods in both of the distinguished processes. 
	An account that we suggest is grounded in the notion of affective loop, which in our proposal is realized by two components: affective game design patterns, and affective physiological reaction patterns of the player. 
	We satisfy the current demand for ubiquitousness of the AfC hardware by focusing on %easily accessible,
        wearable sensors.
%%%	Moreover, we assume applicability of our proposition for designing computer systems in general, not only in the context of video games.
%% niby jak???
	
	The rest of the paper is organized as follows. 
	In Sect.~\ref{sec:afloop} we explain how the affective loop is realized in video games, which is our research ground of choice. 
	Sect.~\ref{sec:afgdp} provides a~deeper insight into the cornerstone of our suggested approach -- the affective game design patterns. 
	Next, in Sect.~\ref{sec:exps}, a description of the experiments conducted for verification of our hypotheses so far is presented. %, as well as our experimental procedure and current result evaluation. 
	The paper is concluded with an overview of selected related studies in Sect.~\ref{sec:related}, and also conclusions and future research directions in Sect.~\ref{sec:end}.
	
	\section{Affective Loop and Games}
	\label{sec:afloop}
	
	One could say that affective gaming is a Holy Grail of affective computing. 
	This is because it realizes the \textit{affective loop}~\cite{hook2008affective} in the most clear and apparent way. 
	The affective game engine reacts in real-time to the player's (affective) input, by detecting relevant data, interpreting it, and generating response by means of modifying various game parameters. 
	This, in turn, creates new circumstances for the player, to which new emotional reactions may appear, and the loop continues. 
	The changes considered may appear on the level of mechanics, but also of dynamics and even aesthetics~\cite{hunicke2004mda}. 
	Bringing affective dimension into games elevates the desired designer's intentions -- be it player's satisfaction or % other factors, such as
        educational or therapeutic impact. 
	
	In comparison to other %types of
        software, games by their nature are an extremely interactive medium. 
	It is common for them to engage in a specific type of communication with the player. 
	This kind of interplay is often evaluated in terms of \textit{immersiveness}, which refers to the perceived engagement in the challenge that the player willingly takes~\cite{Suits2005}. 
	Usually it is directly connected to the notion of flow~\cite{csikszentmihalyi1990flow}, a state where the person stays focused on the performed task while also feeling enjoyed and motivated. 
	This is precisely what is addressed by the affective loop. % implementation. 
	
	%An effective affective loop, however, needs reliable player affect models and decent game engine that cover emotional layers of gameplay. 
	Affective loop~\cite{hook2008affective}, although being a novel and innovative idea as it is, can use much more improvement. 
	As it has been noted before, affects may arise or emerge as a~consequence of many different levels of interaction (from mechanics, through dynamics, to aesthetics). 
	Nowadays, emotion detection and modeling can be supported by Artificial Intelligence (AI) methods, which can contribute to better affect recognition and interpretation. 
	This, however, should be considered early in the design phase. 
	
	As such, we suggest a new, improved approach to video game and serious game design. 
	Our aim is to create an environment, where NPC affect models and player affect models converge. 
	To sustain recent standards for pervasive and ubiquitous computing, we focus on easily accessible, off-the-shelf wearable sensors, such as wristbands, as affective data acquisition hardware. 
	Moreover, we recognize the importance of carefully conducted design phase in the context of gaming. 
	To achieve this, we propose a design concept based on affective game design patterns. 
	
	\section{Affective Game Design Patterns Framework}
	\label{sec:afgdp}
	
\begin{figure}
		\centering
		\includegraphics[width=.4\textwidth]{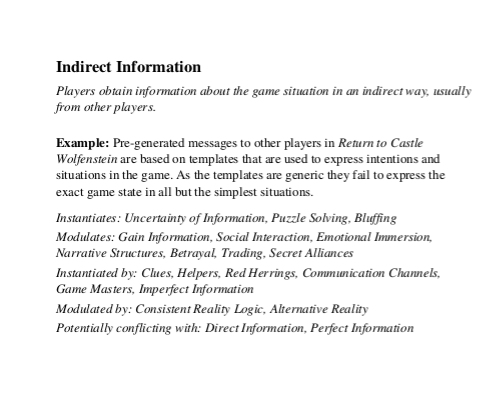}
		\caption{An exemplary Game Design Pattern template, one of several that we consider in our research}
		\label{fig:pat}
	\end{figure}

	Some solutions and mechanics in games tend to reappear in many different games, even across various genres. 
	These include, for example, the action of \textit{Traversing} through the game world or through game levels, \textit{Collecting} some in-game objects (\textit{Pick-ups}), providing the player with \textit{Perfect} (or \textit{Imperfect}) \textit{Information} about game states, \textit{Cooperation} or \textit{Competition} as general playing strategy, and many more. 
	In a~framework proposed by \cite{bjork2005gamepatterns}, such pervasive mechanisms, game design patterns, can be distinguished and used by game developers and researchers alike. 
	Besides having such a collection of patterns to recognize different modes and motifs, the benefit comes also from the patterns forming a hierarchy, and complex net of relationships. 
	The patterns may instantiate or modulate each other, or actually suppress other patterns from appearing (see Figure~\ref{fig:pat}). 
	This way, a better, emergent design of the game can be achieved. 
	
		Research in emotion~\cite{fontaine2007world} suggests that several dimensions characteristic for these states can be distinguished. 
	From the affective computing point of view, a~more classic approach, taking into account mostly two aspects, is considered to be useful. 
	These qualities are arousal and valence~\cite{mauss2009measures}. 
	Arousal distinguishes the level of activity of a given state and valence is associated with pleasure or lack of it~\cite{russell1980circumplex}. 
	Biological connection to these dimensions can be found in reactions related to the Autonomic Nervous System (ANS). 
	Especially, one is able to measure them using wearable devices recording Heart Rate (HR) and Skin Conductance/Galvanic Skin Response (GSR) levels~\cite{cacioppo2000psychophysiology}.

	Our motivation is the belief that some of the game design patterns by their nature evoke emotional reactions of the player's ANS. 
	Similarly to other patterns, affective ones can occur on many different levels of gameplay, from mechanics and control modes to game world aesthetics and the interface. % features. 
	We focus on more basic levels of interaction, and suggest that simple game events caused by including those patterns in game design will meet player's affective response (i.~e.~stress induced by \textit{Time Limit}, or appearance of \textit{Enemies}). 
	Player's emotion elicitation, in turn, can be observed on the level of physiological signals, including Heart Rate (HR) and Galvanic Skin Response (GSR). 
	By means of correlating the affective game design patterns with the biological reaction patterns, we propose a new approach to game design, which integrates both emotion detection and emotion modeling. 
	
\begin{figure}[!ht]
          \centering
        \includegraphics[width=0.38\textwidth]{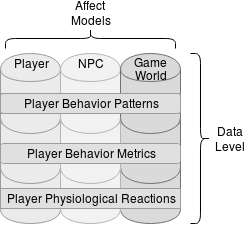}
        \caption{A hierarchy of affective data gathering levels encompassing all affective game modeling perspectives}
         \label{fig:hier}
        \end{figure}

	In our proposed framework, the developer has a better understanding of the gaming experience in the design phase, see Figure~\ref{fig:hier}. 
	This is enabled by access to information on the affective nature of various game design patterns, supported by physiological emotional reaction patterns displayed by the player. 
	Knowing which game elements are evoking affective responses, and confronting this with how actually the player reacts to these elements, allows for better realization of the affective loop mechanism. 
	Additionally, as the affective loop relies greatly on efficient processing of large amounts of data and providing convincingly responsive game environment, AI methods can provide just that. 
	%still nothing about WHAT methods exactly we will use and HOW will they influence the system...

        % FIXME BGC PHD
        We are targeting the Unity game engine as our research ground, although there are other alternatives.
        We are aiming at supporting game designers by a catalog of game design patterns, with the focus of the affective ones.
        The cataloger will be semantically annotated using the ontology of patters that we are creating.
        During the design, a recommendation module will help the designer to select proper patterns, as well as relevant alternatives.
        Furthermore, we are developing an emotion detection layer, currently using the HR and GSR signals.
        Based on them, an emotion classification layer will be provided.
        We are currently evaluating several classification techniques, basing on the work described in~\cite{DBLP:conf/softcomp/RinconCNJC16}.
        The emotion interpretation layer will be provided for the game developer to offer basic reasoning about the emotional state of the player.
        We are aiming to use ontological reasoning~\cite{berthelon2013emoca}. 
        At this level, it will be possible to connect the symbolic model of emotion of the player with some of the existing models of emotions of NPCs~\cite{gonzales2011abe}
.        
	In order to verify our account, we provide empirical studies in the next section.
	
	\section{Experiments}
	\label{sec:exps}

	The experiments described in this section only evaluate our initial assumptions regarding the emotion detection layer, as well as the gathering of data for the classification layer.
	Furthermore, we mention two practical game prototypes demonstrating the affective loop.
        
	\subsection{Outline of procedure}
	
	%Using present research,
        We prepared an examination consisting eventually of three phases. 
	In the first stage, called Calibration Phase, the participant was presented with affective pictures~\cite{marchewka2014naps} for a fixed amount of time. 
	The task was based on subjective evaluation of arousal evoked by every picture. 
	Application was generated using PsychoPy Builder interface and then reprocessed in Python language. 
	At the same time, the participant of the study wears devices that record HR and GSR. 
	The goal of this phase was to create physiological patterns in response to the presented stimuli. 
	Their preparation was to allow verification of hypothesis regarding the impact of affective game design patterns, and to allow development of applications with affective loop implemented in the program in the future works of other research~\cite{nalepa2017fedcsis}.
	
	%\begin{figure}[!ht]
	%\centering
	%\includegraphics{eHset1}
	%\caption{Participant's hand connected to e-Health HR and GSR sensors, while wearing Empatica E4 wristband.}
	%\label{fig:setup}
	%\end{figure}
	
\begin{figure}[h!]
		\centering
		\includegraphics[width=0.4\textwidth]{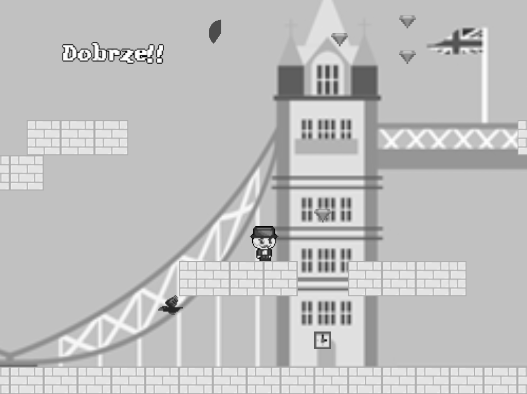}
		\caption{A screenshot from the gameplay of ``London Bridge'', a~game designed specifically for this research purposes.}
		\label{fig:game}
	\end{figure}

	Next phase was called Gaming Phase and the subject's task was to play an affective computer game. 
	It was a platform side-scrolling game, designed using affective game design patterns as adapted from another study~\cite{nalepa2017fedcsis} (see Figure~\ref{fig:game}). The picture shows some of the patterns that we consider to have affective impact: Time Limit (a pie chart depicting time left to complete the level), Indirect Information (time and score being represented using a chart and a description, not explicitly indicated numerically, compare with Figure~\ref{fig:pat}) and Enemies (a Crow on the bottom-left part of the Figure~\ref{fig:game}).
        
	The whole program was created using GameMaker environment. 
	The main task of the player was to navigate the given space in order to get as many points as possible in a~given time. 
	As an impediment, we used traps and opponents which disturbed the participant. 
	Devices measuring physiological signals like HR and GSR were also used in this stage of experiment. 
	
	In the last phase, the task of the subject was to watch presented neutral (in terms of valence) picture, which was chosen from NAPS database~\cite{marchewka2014naps}. 
	After fixed time, the sound material was presented to the subject, a female cry with a strong affective impact. 
	The reason for including this step was to acquire participant's readings of a~strong reaction to an affective stimuli, to be used in further pattern development. 
	The wearable devices were used once again.
	
	\subsection{Used platforms}
	
	Physiological signals were measured using several devices. 
	Two of them, BITalino and eHealth, are extendable platforms with boards and sensors coming extra. 
	Others, namely Empatica E4 and Microsoft Band 2, are dedicated for more common usage. 
	Along with sensors, a PC and a smartphone were used in our research.
	
	Data from BITalino and eHealth was collected using standard Bluetooth and USB interfaces. 
	The whole mechanism of acquisition was written in Python language. 
	From the scientific point of view, the method of acquiring of physiological signals from wristbands is more interesting. 
	We decided to prepare our own solution -- BandReader -- an Android application (for more information on this application see~\cite{bandreader-icaisc2018}). 
	It enables the user to collect data from several devices simultaneously. 
	
	\subsection{Experiments summary}
	
	During our scientific research we conducted series of experiments, which consisted of different elements.
	First examination was carried out on 6 people and only with the Calibration phase. 
	This experiment made it possible to test the created application. 
	Gaming Phase was added in another session of tests. 
	This time, 9 people were examined. 
	The results of this study have been published elsewhere, in~\cite{nalepa2017fedcsis}.
	
	In November and January, we conducted two studies to determine whether the devices used by us are suitable for our purposes. 
	We used Neurobit device as a reference for HR and GSR, and also Polar Pulse chest strap for HR. 
	21 subjects participated. 
	As in the previous tests, we applied first two phases. 
        Experiments in November and January helped us to determine which devices ought to be used in our future work.

        Comparison of data from wearable devices and from neuromedical references helped us to determine devices that are highly promising in research. 
	We managed to establish that eHealth correctly measures HR, and Empatica E4 can be useful to collect GSR signals. 
	Unfortunately, Microsoft Band 2 proved to be unreliable as a data acquisition device for our needs.
	It turned out that data from BITalino is highly accurate both in terms of HR and GSR signals. 
        As such, we decided to focus on BITalino.
        Furthermore, it is a well supported and low-cost platform.
        
%	In order to collect as much data as possible, we abandon the assumption that every participant of the examination has to be hooked to all devices. 
%	The reason being simply technical, meaning to maximize the measuring potential of each of the platforms. 
%	For this, we prepared 3 separate stations with different devices. 
%	The first one was consisted of BITalino, in the second we used eHealth and Empatica, and the last one was a~combination of former two. 

	The last session of experiments was conducted in late March 2018. 
	The goal was to collect data that could be used to create models for emotion detection and classification, enabling the analysis of emotional states.
        As we opted for BITalino, in this session we used only this single measurement device.
	We adopted 3 phases of the experiment to acquire data, which contain as many information about emotional states as possible. 
	We managed to examine roughly 100 people in this way.
        Currently we are still analyzing the data, to build the emotion classification layer.

        As it was mentioned, in first experiments we used a prototype game developed with GameMaker. % Studio software before, which proved to be a~decent tool for game creation, but not sufficient for affectively adaptive games.
	However, to address the industry standards, %For these reasons, to adress this chalange
        we are now experimenting with our own game prototypes in Unity. %and tutorials.
%%        This is related to the cooperation of the AGH UST and the Jagiellonian university students,
	In particular, together with Kamil Osuch from AGH UST, in May 2018 we developed a~simple Asteroids game in Unity coupled with BITalino sensors for implementing affective loop. 
	This prototype could be presented during the conference in a form of a playable demo.

%	\subsection{Evaluation}

	\section{Related Works}
	\label{sec:related}
	
	It is evident that the opportunity for AI to support affect models (including the gaming context) has already been recognized.  
	This section provides an overview of selected other related research in this area. 
	
	The general AI uses in games in recent years tend to cover four main aspects: player experience models, procedural generation of the content, massive-scale player data mining and enhancing NPC capabilities~\cite{yannakakis2012gameai}. 
	However, this does not seem to be reflected in the advances on affective realism in games, at least not apparently~\cite{hudlicka2009affective}. %IDEA FOR A PAPER: overview of affective game ai state-of-art
	Nevertheless, different researchers engaged in studies on the use of, for example, machine learning for example for emotion recognition~\cite{becker2005physiologically,sabourin2011modeling,rani2006empirical,shang2017continuous}, 
	emotion modeling for providing predictions~\cite{conati2002modeling,zhou2003inferring,camilleri2017towards,shang2017continuous}, for engaging Dynamic Difficulty Adjustment (DDA)~\cite{rani2005maintaining,liu2009dynamic}, 
	or considering other aspects of game, such as camera control~\cite{yannakakis2010towards}. 
	
	Meanwhile, other existing frameworks for affective game design do not seem to reach for AI methods at all. 
	\cite{kolakowska2013emotion} and \cite{szwoch2016evaluation} propose specific approach towards affective games design process. 
	\cite{dormann2013once}, \cite{caminha2017development} and \cite{nalepa2017fedcsis} suggest using game design patterns and reappearing game mechanics for development of emotional layer of games.
	%FIXED:
	Neither of those accounts refers to facilitation using AI methods directly. 
	
	\section{Conclusion and Future Plans}
	\label{sec:end}
	
	The key theme of this paper is to highlight an opportunity for AI to facilitate affective dimension of human-computer interaction. 
	Emotion detection and emotion modeling, when supported by AI methods, can benefit from enhanced accuracy, realism and reliability. 
	However, we believe that both of these aspects should be considered early in the design phase of the system. 
	We choose video game design as our research ground, where the affective loop mechanism can be realized to the greatest extent. 
	We propose a new approach to affective games design, along with a suggestion that it can be expanded to encompass all AI dependent types of software. 
	
	Our concept signifies the game design phase as the best moment for introducing the affective loop, thus providing a~meeting point of affective data collection and affect models. 
	Both stages of the loop are enhanced by AI techniques. 
	Furthermore, the data collection is included in the loop by means of affective physiological reaction patterns of the player. 
	It is an essential assumption that those emotional responses are in close relation to affective design patterns that underlie the game design. 
%%%	The correlations between affective reactions and affective game events form a keystone of our approach. 
%% tja, ale ich nie mamy!
	
	The future directions of the described research include further and deeper analysis of data acquired during conducted experiments. 
	Specifically, a search of correlations between affective design patterns and affective physiological reaction patterns is anticipated. 
	While we also consider a shift towards another game environment (from GameMaker to Unity), we aim at introducing proper affective loop in the experimental game design as well. 
	
	Another issue that will be addressed is the hardware setup. 
	While the quality of raw data acquired is now acceptable, there is still room for improvement regarding the unobtrusiveness requirement for the used devices. 
	As Empatica wristband is quite acceptable in this matter, BITalino and eHealth electrodes have to be reconsidered in terms of, for example, packing them into some 3D printed wearable devices. % or comfortable game controllers. 

        We are also working on the integration of BITalino readings with the Unity game engine. % to experiment with simple affective loop implementations.
	To this goal we are using the Unity API developed for BITalino. 
	At the moment, we are exploring how the Unity developed game mechanics interact with real-life data from the player using our hardware setup.   
	
%	As for our future plans, we will build on current efforts indicated in~\ref{subsec:chal}. 
%	We intend to use acquired data to find correlations between affective game events brought by affective design patterns and physiological reactions of the player. 
%	This set of patterns will then serve for further tests of our framework, most likely in a game that will include an affective loop mechanism. 
%	The outcomes of this undertaking will be presented at the conference. 

%	As for the data handling, we intend to further extend and improve BandReader, our smartphone data recording software. 
%	Support of more wristband models is considered, as well as implementing a better synchronization layer and a~broadcasting mechanism. 
	
%	Possible applications of emotionally intelligent technologies are vast. 
%	Although the development of new AI techniques and affective models is quite vigorous, it seems that currently there is a lack of comprehensive design frameworks in this context. 
%	We expect that our research will contribute significantly to clearer and better insight for providing precise emotion detection and solid emotion modeling. 
	

        \section*{Acknowledgments}
        \label{sec:acknowledgements}
        The paper is supported by the AGH University research grant.

	\bibliographystyle{named}
	\bibliography{afcaipub,geistbib/geistpub,geistbib/geisteam,afcbib/afcaiteam,ijcai18}

\begin{thebibliography}{}

\bibitem[\protect\citeauthoryear{Becker \bgroup \em et al.\egroup
  }{2005}]{becker2005physiologically}
Christian Becker, Arturo Nakasone, Helmut Prendinger, Mitsuro Ishizuka, and
  Ipke Wachsmuth.
\newblock Physiologically interactive gaming with the 3d agent max.
\newblock 2005.

\bibitem[\protect\citeauthoryear{Berthelon and
  Sander}{2013}]{berthelon2013emoca}
F.~Berthelon and P.~Sander.
\newblock Emotion ontology for context awareness.
\newblock In {\em 2013 IEEE 4th International Conference on Cognitive
  Infocommunications (CogInfoCom)}, pages 59--64, Dec 2013.

\bibitem[\protect\citeauthoryear{Bj\"{o}rk and
  Holopainen}{2005}]{bjork2005gamepatterns}
Staffan Bj\"{o}rk and Jussi Holopainen.
\newblock {\em Patterns in Game Design}.
\newblock Charles River Media, 2005.

\bibitem[\protect\citeauthoryear{Cacioppo \bgroup \em et al.\egroup
  }{2000}]{cacioppo2000psychophysiology}
John~T. Cacioppo, Gary~G. Berntson, Jeff~T. Larsen, Kirsten~M. Poehlmann, and
  Tiffany~A. Ito.
\newblock The psychophysiology of emotion.
\newblock In {\em Handbook of emotions}, pages 173--191. Guildford Press, 2000.

\bibitem[\protect\citeauthoryear{Camilleri \bgroup \em et al.\egroup
  }{2017}]{camilleri2017towards}
Elizabeth Camilleri, Georgios~N Yannakakis, and Antonios Liapis.
\newblock Towards general models of player affect.
\newblock In {\em Affective Computing and Intelligent Interaction (ACII), 2017
  International Conference on}, 2017.

\bibitem[\protect\citeauthoryear{Caminha}{2017}]{caminha2017development}
David Capelo~Chaves Caminha.
\newblock Development of emotional game mechanics through the use of biometric
  sensors.
\newblock 2017.

\bibitem[\protect\citeauthoryear{Conati and Zhou}{2002}]{conati2002modeling}
Cristina Conati and Xiaoming Zhou.
\newblock Modeling students’ emotions from cognitive appraisal in educational
  games.
\newblock In {\em International Conference on Intelligent Tutoring Systems},
  pages 944--954. Springer, 2002.

\bibitem[\protect\citeauthoryear{Csikszentmihalyi}{1990}]{csikszentmihalyi1990flow}
Mihaly Csikszentmihalyi.
\newblock Flow: The psychology of optimal performance.
\newblock {\em NY: Cambridge UniversityPress}, 40, 1990.

\bibitem[\protect\citeauthoryear{Dormann \bgroup \em et al.\egroup
  }{2013}]{dormann2013once}
Claire Dormann, Jennifer~R Whitson, and Max Neuvians.
\newblock Once more with feeling: Game design patterns for learning in the
  affective domain.
\newblock {\em Games and Culture}, 8(4):215--237, 2013.

\bibitem[\protect\citeauthoryear{Fontaine \bgroup \em et al.\egroup
  }{2007}]{fontaine2007world}
Johnny~RJ Fontaine, Klaus~R Scherer, Etienne~B Roesch, and Phoebe~C Ellsworth.
\newblock The world of emotions is not two-dimensional.
\newblock {\em Psychological science}, 18(12):1050--1057, 2007.

\bibitem[\protect\citeauthoryear{Gi\.zycka and Nalepa}{2018}]{bgc2018gem}
Barbara Gi\.zycka and Grzegorz~J. Nalepa.
\newblock Emotion in models meets emotion in design: building true affective
  games.
\newblock submitted to IEEE GEM 2018, 2018.

\bibitem[\protect\citeauthoryear{Gonzalez-Sanchez \bgroup \em et al.\egroup
  }{2011}]{gonzales2011abe}
J.~Gonzalez-Sanchez, M.~E. Chavez-Echeagaray, R.~Atkinson, and W.~Burleson.
\newblock Abe: An agent-based software architecture for a multimodal emotion
  recognition framework.
\newblock In {\em 2011 Ninth Working IEEE/IFIP Conference on Software
  Architecture}, pages 187--193, June 2011.

\bibitem[\protect\citeauthoryear{H{\"o}{\"o}k}{2008}]{hook2008affective}
Kristina H{\"o}{\"o}k.
\newblock Affective loop experiences--what are they?
\newblock In {\em International Conference on Persuasive Technology}, pages
  1--12. Springer, 2008.

\bibitem[\protect\citeauthoryear{Hudlicka}{2009}]{hudlicka2009affective}
Eva Hudlicka.
\newblock Affective game engines: motivation and requirements.
\newblock In {\em Proceedings of the 4th international conference on
  foundations of digital games}, pages 299--306. ACM, 2009.

\bibitem[\protect\citeauthoryear{Hunicke \bgroup \em et al.\egroup
  }{2004}]{hunicke2004mda}
Robin Hunicke, Marc LeBlanc, and Robert Zubek.
\newblock Mda: A formal approach to game design and game research.
\newblock In {\em Proceedings of the AAAI Workshop on Challenges in Game AI},
  volume~4, pages 1--5. AAAI Press San Jose, CA, 2004.

\bibitem[\protect\citeauthoryear{Ko{\l}akowska \bgroup \em et al.\egroup
  }{2013}]{kolakowska2013emotion}
Agata Ko{\l}akowska, Agnieszka Landowska, Mariusz Szwoch, Wioleta Szwoch, and
  Micha{\l}~R Wr{\'o}bel.
\newblock Emotion recognition and its application in software engineering.
\newblock In {\em Human System Interaction (HSI), 2013 The 6th International
  Conference on}, pages 532--539. IEEE, 2013.

\bibitem[\protect\citeauthoryear{Koller and Friedman}{2009}]{KollerFriedman:09}
Daphne Koller and Nir Friedman.
\newblock {\em Probabilistic Graphical Models: Principles and Techniques}.
\newblock MIT Press, 2009.

\bibitem[\protect\citeauthoryear{Kutt \bgroup \em et al.\egroup
  }{}]{bandreader-icaisc2018}
Krzysztof Kutt, Grzegorz~J. Nalepa, Barbara Gi\.zycka, Pawe\l{} Jemio\l{}o, and
  Marcin Adamczyk.
\newblock Bandreader -- a mobile application for data acquisition from wearable
  devices in affective computing experiments.
\newblock submitted to ICAISC 2018.

\bibitem[\protect\citeauthoryear{Liu \bgroup \em et al.\egroup
  }{2009}]{liu2009dynamic}
Changchun Liu, Pramila Agrawal, Nilanjan Sarkar, and Shuo Chen.
\newblock Dynamic difficulty adjustment in computer games through real-time
  anxiety-based affective feedback.
\newblock {\em International Journal of Human-Computer Interaction},
  25(6):506--529, 2009.

\bibitem[\protect\citeauthoryear{Marchewka \bgroup \em et al.\egroup
  }{2014}]{marchewka2014naps}
Artur Marchewka, {\L}ukasz {\.{Z}}urawski, Katarzyna Jednor{\'o}g, and Anna
  Grabowska.
\newblock The {N}encki {A}ffective {P}icture {S}ystem ({NAPS}): Introduction to
  a novel, standardized, wide-range, high-quality, realistic picture database.
\newblock {\em Behavior Research Methods}, 46(2):596--610, 2014.

\bibitem[\protect\citeauthoryear{Mauss and Robinson}{2009}]{mauss2009measures}
Iris~B. Mauss and Michael~D. Robinson.
\newblock Measures of emotion: A review.
\newblock {\em Cognition and Emotion}, 23(2):209--237, 2009.

\bibitem[\protect\citeauthoryear{Nalepa \bgroup \em et al.\egroup
  }{2017}]{nalepa2017fedcsis}
Grzegorz~J. Nalepa, Barbara Gizycka, Krzysztof Kutt, and Jan~K. Argasinski.
\newblock Affective design patterns in computer games. scrollrunner case study.
\newblock In {\em Communication Papers of the 2017 Federated Conference on
  Computer Science and Information Systems, FedCSIS 2017}, pages 345--352,
  2017.

\bibitem[\protect\citeauthoryear{Nalepa \bgroup \em et al.\egroup
  }{2018}]{gjn2017fgcs}
Grzegorz~J. Nalepa, Krzysztof Kutt, and Szymon Bobek.
\newblock Mobile platform for affective context-aware systems.
\newblock {\em Future Generation Computer Systems}, 2018.

\bibitem[\protect\citeauthoryear{Picard}{1997}]{picard1997affective}
Rosalind~W. Picard.
\newblock {\em Affective Computing}.
\newblock MIT Press, 1997.

\bibitem[\protect\citeauthoryear{Rani \bgroup \em et al.\egroup
  }{2005}]{rani2005maintaining}
Pramila Rani, Nilanjan Sarkar, and Changchun Liu.
\newblock Maintaining optimal challenge in computer games through real-time
  physiological feedback.
\newblock In {\em Proceedings of the 11th international conference on human
  computer interaction}, volume~58, pages 22--27, 2005.

\bibitem[\protect\citeauthoryear{Rani \bgroup \em et al.\egroup
  }{2006}]{rani2006empirical}
Pramila Rani, Changchun Liu, Nilanjan Sarkar, and Eric Vanman.
\newblock An empirical study of machine learning techniques for affect
  recognition in human--robot interaction.
\newblock {\em Pattern Analysis and Applications}, 9(1):58--69, 2006.

\bibitem[\protect\citeauthoryear{Rincon \bgroup \em et al.\egroup
  }{2016}]{DBLP:conf/softcomp/RinconCNJC16}
Jaime~Andres Rincon, {\^{A}}ngelo Costa, Paulo Novais, Vicente Juli{\'{a}}n,
  and Carlos Carrascosa.
\newblock Using non-invasive wearables for detecting emotions with intelligent
  agents.
\newblock In Manuel Gra{\~{n}}a, Jos{\'{e}}~Manuel L{\'{o}}pez{-}Guede, Oier
  Etxaniz, {\'{A}}lvaro Herrero, H{\'{e}}ctor Quinti{\'{a}}n, and Emilio
  Corchado, editors, {\em International Joint Conference
  SOCO'16-CISIS'16-ICEUTE'16 - San Sebasti{\'{a}}n, Spain, October 19th-21st,
  2016, Proceedings}, volume 527 of {\em Advances in Intelligent Systems and
  Computing}, pages 73--84, 2016.

\bibitem[\protect\citeauthoryear{Russell}{1980}]{russell1980circumplex}
J.~A. Russell.
\newblock A circumplex model of affect.
\newblock {\em Journal of Personality and Social Psychology}, 39(6):1161--1178,
  1980.

\bibitem[\protect\citeauthoryear{Sabourin \bgroup \em et al.\egroup
  }{2011}]{sabourin2011modeling}
Jennifer Sabourin, Bradford Mott, and James~C Lester.
\newblock Modeling learner affect with theoretically grounded dynamic bayesian
  networks.
\newblock In {\em International Conference on Affective Computing and
  Intelligent Interaction}, pages 286--295. Springer, 2011.

\bibitem[\protect\citeauthoryear{Shang}{2017}]{shang2017continuous}
Zhengkun Shang.
\newblock Continuous affect recognition with different features and modeling
  approaches in evaluation-potency-activity space.
\newblock Master's thesis, University of Waterloo, 2017.

\bibitem[\protect\citeauthoryear{Suits}{2005}]{Suits2005}
B.~Suits.
\newblock {\em The Grasshopper: Games, Life and Utopia}.
\newblock Broadview Press, 2005.

\bibitem[\protect\citeauthoryear{Szwoch}{2016}]{szwoch2016evaluation}
Mariusz Szwoch.
\newblock Evaluation of affective intervention process in development of
  affect-aware educational video games.
\newblock In {\em Computer Science and Information Systems (FedCSIS), 2016
  Federated Conference on}, pages 1675--1679. IEEE, 2016.

\bibitem[\protect\citeauthoryear{Thompson and
  McGill}{2015}]{thompson2015affhci}
Nik Thompson and Tanya McGill.
\newblock Affective human-computer interaction.
\newblock In {\em Encyclopedia of Information Science and Technology, Third
  Edition}, pages 3712--3720. IGI Global, 2015.

\bibitem[\protect\citeauthoryear{Yannakakis \bgroup \em et al.\egroup
  }{2010}]{yannakakis2010towards}
Georgios~N Yannakakis, H{\'e}ctor~P Mart{\'\i}nez, and Arnav Jhala.
\newblock Towards affective camera control in games.
\newblock {\em User Modeling and User-Adapted Interaction}, 20(4):313--340,
  2010.

\bibitem[\protect\citeauthoryear{Yannakakis}{2012}]{yannakakis2012gameai}
Geogios~N Yannakakis.
\newblock Game ai revisited.
\newblock In {\em Proceedings of the 9th conference on Computing Frontiers},
  pages 285--292. ACM, 2012.

\bibitem[\protect\citeauthoryear{Zhou and Conati}{2003}]{zhou2003inferring}
Xiaoming Zhou and Cristina Conati.
\newblock Inferring user goals from personality and behavior in a causal model
  of user affect.
\newblock In {\em Proceedings of the 8th international conference on
  Intelligent user interfaces}, pages 211--218. ACM, 2003.

\end{thebibliography}
	
\end{document}